# On the Persistent Shape and Coherence of Pulsating Auroral Patches

B. K. Humberset[1], J. W. Gjerloev[1,2], I. R. Mann[3], R. G. Michell[4,5], and M. Samara[5]

[1]Birkeland Centre for Space Science, Department of Physics and Technology, University of Bergen, Bergen, Norway, [2]The Johns Hopkins University Applied Physics Laboratory, Laurel, MD, USA, [3]Department of Physics, University of Alberta, Edmonton, Alberta, Canada, [4]Department of Astronomy, University of Maryland, College Park, MD, USA, [5]NASA Goddard Space Flight Center, Greenbelt, MD, USA

**Key Points:**
- The shapes of the patches are remarkably persistent over periods of 4.5–8.5 min
- Only one of four patches fluctuates in a coherent fashion
- The patches appear to drift differently from the SuperDARN-determined ExB velocity

**Supporting Information:**
- Supporting Information S1
- Movie S1
- Movie S2
- Movie S3
- Movie S4
- Movie S5
- Movie S6

**Correspondence to:**
B. K. Humberset,
beate.humberset@uib.no

**Abstract** The pulsating aurora covers a broad range of fluctuating shapes that are poorly characterized. The purpose of this paper is therefore to provide objective and quantitative measures of the extent to which pulsating auroral patches maintain their shape, drift and fluctuate in a coherent fashion. We present results from a careful analysis of pulsating auroral patches using all-sky cameras. We have identified four well-defined individual patches that we follow in the patch frame of reference. In this way we avoid the space-time ambiguity which complicates rocket and satellite measurements. We find that the shape of the patches is remarkably persistent with 85–100% of the patch being repeated for 4.5–8.5 min. Each of the three largest patches has a temporal correlation with a negative dependence on distance, and thus does not fluctuate in a coherent fashion. A time-delayed response within the patches indicates that the so-called streaming mode might explain the incoherency. The patches appear to drift differently from the SuperDARN-determined $\vec{E} \times \vec{B}$ convection velocity. However, in a nonrotating reference frame the patches drift with 230–287 m/s in a north eastward direction, which is what typically could be expected for the convection return flow.

## 1. Introduction

The broad definition of pulsating aurora covers low-intensity aurora that undergoes repetitive, quasiperiodic, or occasionally periodic intensity fluctuations on time scales ranging from less than 1 s to several tens of seconds (Royrvik & Davis, 1977).

### 1.1. Highly Variable Characteristics

The pulsating aurorae occur in very different shapes such as east-west aligned bands, arc segments, and patches of irregular shapes with horizontal sizes of a few to hundreds of kilometers. They do however have joint characteristics other than their intensity fluctuations. For example, they are associated with time-varying energetic electron precipitation (e.g., Evans et al., 1987; Jaynes et al., 2013; Samara et al., 2015). The energy distributions are within a few keV to several tens of keV, but also energies of up to 140 keV (Sandahl et al., 1980) and cases of surprisingly low energies (1–2 keV; McEwen et al., 1981) are reported. Together with a wide range of altitude measurements of the auroral emissions (e.g., Hosokawa & Ogawa, 2015; Jones et al., 2009), this suggests that the electron energies vary from one event to the next and also within a train of fluctuations. Thus, the presumably joint characteristics exhibit a tremendous variation and might not be a meaningful defining characteristic. It is a mystery whether pulsating auroras can have different mechanisms as suggested by Sato et al. (2004), or whether the large variation can be explained by one mechanism. To understand the latter we also need to solve some fundamental questions on pulsating aurora, such as what underlying mechanism(s) controls their on-off fluctuation, shape, and coherency. This illustrates the need for better characterization. We therefore do a careful study of the shape, coherency, and drift of pulsating aurora in the form of patches.

### 1.2. Persistence of Pulsating Patches

While some pulsating auroras only repeat for a few times, a common characteristic of pulsating patches is their persistence in shape. The shape, including intricate detail, persist for many pulsations and only changes on a time scale of minutes (Johnstone, 1978). In fact, as Humberset et al. (2016) pointed out, the characteristics of pulsating patches are erratic in their behavior. The only parameter that seems to be persistent may be their shape, and this could in fact be reviewed as the loose definition of pulsating aurora. However, the underlying reason for this persistence remains unclear (Lessard, 2012). The shapes of the patches have been suggested to be governed by cold plasma of ionospheric origin trapped in flux tubes or ducts (Oguti, 1976), which experience almost no gradient and curvature drift and therefore keep their shape. However, this also assumes that the magnetic field line mapping morphology is maintained between the magnetosphere and the ionosphere. The underlying reason for the persistent shape of pulsating patches as they disappear and reappear repeatedly over several minutes is therefore not fully understood.

### 1.3. Cause of Spatiotemporal Variation Is Still Debated

It is not known to what extent the pulsating auroral shapes fluctuate coherently. Three fundamentally different modes have been proposed (e.g., Kosch & Scourfield, 1992; Royrvik & Davis, 1977; Scourfield & Parsons, 1971; Thomas & Stenbaek-Neelsen, 1981; Yamamoto & Oguti, 1982): (1) the standing/pure mode is a synchronous intensity fluctuation over the entire form, (2) the streaming/expansion mode usually involves outward growth followed by contraction or disappearance as the pulse decays, (3) the propagating/moving mode where a patch brightens quickly and sweeps away from its original position as the intensity starts to drop, or forms that propagate laterally across the sky continuously or sequentially. It is not known why the shapes sometimes fluctuate coherently and sometimes not, and the underlying cause of the spatiotemporal variations remains to be determined.

### 1.4. Patch Drift

One of the outstanding issues associated with fluctuating patches is whether they drift with the $\vec{E} \times \vec{B}$ velocity. Answering this apparently simple question is, however, difficult due to observational complexities. The Super Dual Auroral Radar Network (SuperDARN) provides observations with a spatial resolution of some 50 km by 50 km for the measured line-of-sight backscatter, but the resolution of the final convection solution is far coarser as it requires the inclusion of a statistical fill-in model. Spaceborne measurements are also complicated due to their inability in separating spatial and temporal variations. It takes some 7 s for a low-Earth-orbit (LEO) satellite to traverse a patch of 50 km, which is comparable to the on time of the patch, and thus it is impossible to separate spatial and temporal variations. It should also be mentioned that single-satellite measurements only provide observations along the satellite trajectory and thus the 2-D convection distribution cannot be unambiguously inferred.

It has been suggested that imaging of pulsating auroral patches can be used to remote sense magnetospheric convection (e.g., Nakamura & Oguti, 1987; Yang et al., 2015; 2017). However, the assumption here is that all monitored fluctuating auroral patches move with the plasma convection $\vec{E} \times \vec{B}$ velocity. The drift speed has been consistently measured to be on the order of 1 km/s in the dawn sector, presumably at the $\vec{E} \times \vec{B}$ velocity (Davis, 1978; Scourfield et al., 1983), but there have been studies which found indications that the drift of the pulsating aurora can be different from the plasma convection (Swift & Gurnett, 1973; Wescott et al., 1976). It is therefore important to establish if all patches are drifting solely with the plasma convection $\vec{E} \times \vec{B}$ velocity.

### 1.5. Purpose of the Paper

Humberset et al. (2016) provided the temporal characteristics of six pulsating auroral patches. They found that the patches were highly variable in all measurable parameters and thus suggested the term “fluctuating aurora” instead of the commonly used term “pulsating aurora.” The problem with pulsating is that it implies the patch to vary regularly. None of the proposed mechanisms appears to explain the observational constraints set by these patches in a convincing manner. In this paper we therefore provide spatial characteristics for the same event.

The purpose of this paper is to provide objective and quantitative measurements of the extent to which pulsating auroral patches maintain their morphology and fluctuate in a coherent fashion. Does the patch maintain its shape? Does the entire patch vary with the same fluctuation and phase? Do the patches drift with the plasma convection $\vec{E} \times \vec{B}$ velocity?

We use a ground-based all-sky imager that provided two-dimensional coverage of an electrodynamic parameter, the 557.7 nm emissions. This has the distinct advantage of allowing for a separation of spatial and temporal variations by translating the emissions to a nonrotating reference frame. To be specific, we trace the patch in its frame of reference, assuming that the drift of the patch is in the nonrotating frame of reference. In this way we avoid the space-time ambiguity, which complicates rocket and satellite measurements. Whereas it would take a LEO satellite less than one fluctuation to cross a patch, a rocket takes a few on-off cycles, for which the observed variation is both temporal and spatial. With all-sky imagers we can follow a pulsating auroral patch over its lifetime. In section 2 we describe the data and event; section 3 outlines the technique and methodology; section 4 shows the pulsating auroral patches and their quantitative characteristics; in section 5 we discuss our results; and finally in section 6 we summarize and draw conclusions.

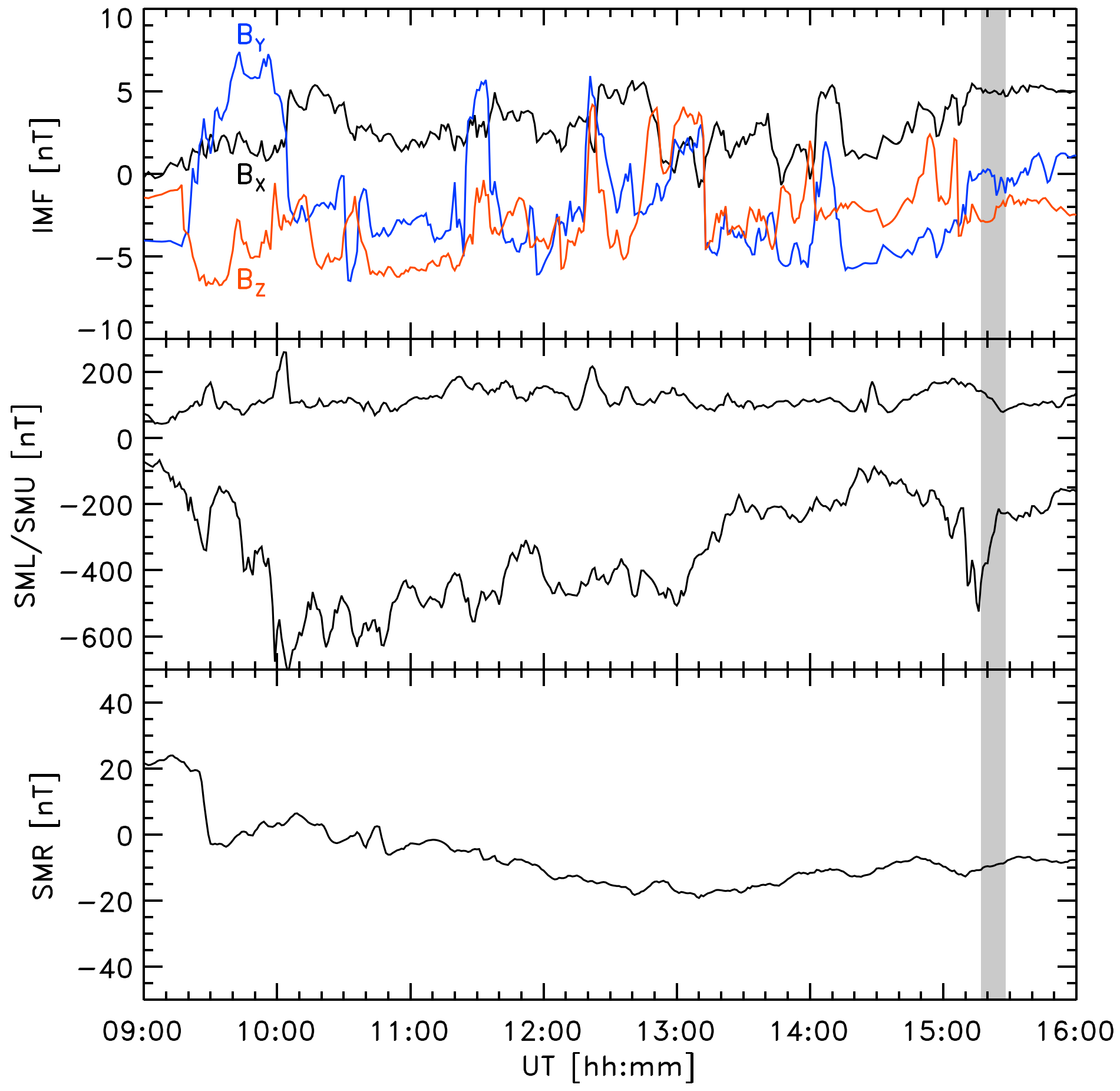

**Figure 1.** Geomagnetic indices (top) IMF, (middle) SML/SMU (SuperMAG equivalent of the AE indices) maximum westward/eastward auroral electrojet strength, and (bottom) SMR (SuperMAG equivalent of the SYMH index) symmetric ring current index. The time of the event is highlighted in gray. We use the SuperMAG data set of indices and ACE IMF data which is propagated to the front of the magnetosphere (courtesy of Dr. James Weygand).

## 2. Data and Event

The all-sky imager (ASI) utilized is located at Poker Flat Research Range in Alaska at −147.4° geographic longitude, 65.1° geographic latitude. The ASI uses a narrow band filter to capture the green line 557.7 nm emissions from atomic oxygen. It produces frames with 512 by 512 pixels resolution at a frame rate of 3.31 Hz.

We utilize the same event and observations as described in Humberset et al. (2016). Their findings and other published results have shown that the pulsating aurora is not regularly periodic, but rather erratic, leading them to suggest that the name "fluctuating aurora" is more appropriate. This is therefore the term used in the following sections. Below we give a short description of the event.

Our event was recorded on 1 March 2012. Figure 1 shows the SuperMAG data set of indices and time-shifted interplanetary magnetic field (IMF; Gjerloev, 2012; Newell & Gjerloev, 2011, 2012). The geomagnetic conditions are moderately disturbed, with a moderately southward IMF (top panel) and a SML index (SuperMAG equivalent of the AL index, middle panel) of the maximum westward auroral electrojet strength showing many hours of almost continuous substorm activity prior to our event. Our event (highlighted in gray) is during the second substorm. Before 15:00 UT the fluctuating aurora covers the southern part of the sky with large east-west structures, which seem to fluctuate in a propagating and/or streaming fashion. At about 15:00 UT the fluctuating aurora starts to break up into more well-defined patches. The four selected patches are located at ~4 magnetic local time (MLT) and ~65° magnetic latitude. Altogether, the data set includes hundreds of fluctuations recorded from about 15:17 to 15:28 UT. The research data are available in the data set Humberset et al. (2018).

## 3. Technique

Prior to the analysis we correct for the distorted fish-eye view of the ASI frames by performing a projection of the emissions onto a Cartesian grid with uniform spatial resolution as described in Humberset et al. (2016).

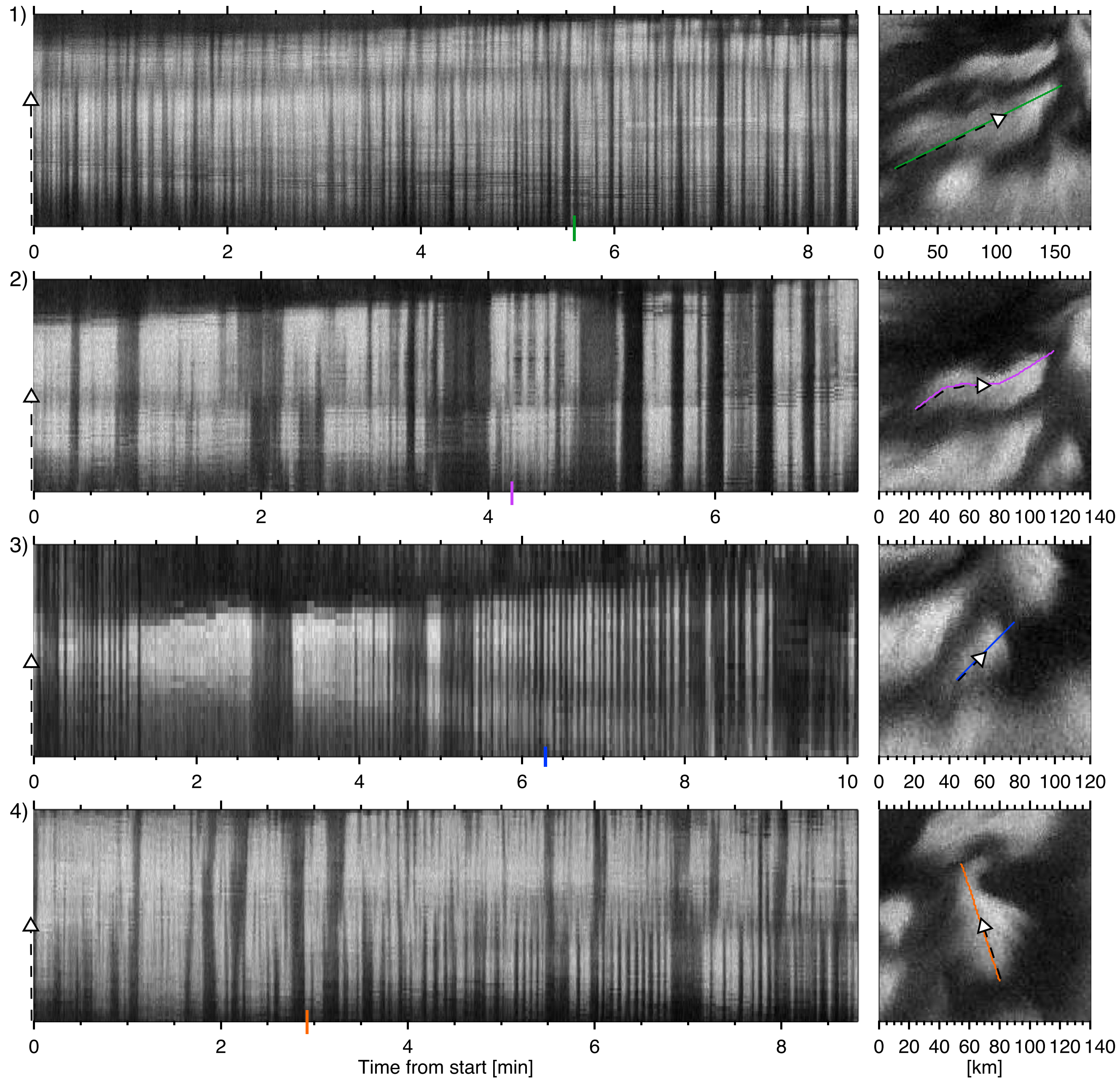

**Figure 2.** For each of the patches (1–4): Evolution, over the time interval analyzed, of a line of pixels crossing the patch (position-time diagrams). The line of pixels are highlighted in color, and the dashed arrows indicate the direction of the sampling. The time of the example image is indicated by a colored tick mark. In step 1 of the technique position-time diagrams are used to decide if the patch is fluctuating and can be considered as one individual and well-defined patch.

We use a thin shell approximation assuming an altitude of 110 km for the auroral emissions. The Cartesian grid has a spatial resolution of 1.0 by 1.0 km and is organized geographically with north on top and west to the right to facilitate a correction of the ASI rotating with the Earth. The most distorted limb pixels are avoided by only including patches that are located within the center 400 by 400 km FOV. To determine the characteristics of the fluctuating auroral patches, we perform the analysis explained in the following subsections.

### 3.1. Patch Identification

The individual patches are manually identified from the movie in the supporting information. However, due to the erratic nature of the fluctuating aurora, this is not straightforward. To determine start and end times of the analysis, we develop position-time diagrams like the ones showed in Figure 2. To make the position-time diagrams, we sample emissions from the same line of pixels for all individual images and plot them successively. The diagrams allow a determination of whether the patch is fluctuating and for how long we can follow it before it either disappears or, for example, joins together with an adjacent patch. More importantly, we use position-time diagrams to decide whether the patch is an individual patch with a different train of fluctuations than the adjacent patches.

Figure 2 shows the evolution for each of the patches (1–4). The line of pixels are highlighted in color and the dashed arrows indicate the direction of the sampling. The time of the example image is indicated by a colored tick mark. The patches are clearly fluctuating in intensity and for each of the patches the variation is convincingly similar across the patch to consider it as an individual patch. However, there is also some intriguing irregularities in the fluctuations across patches 2 and 4. These and the relation to adjacent patches will further be described in section 4.1.

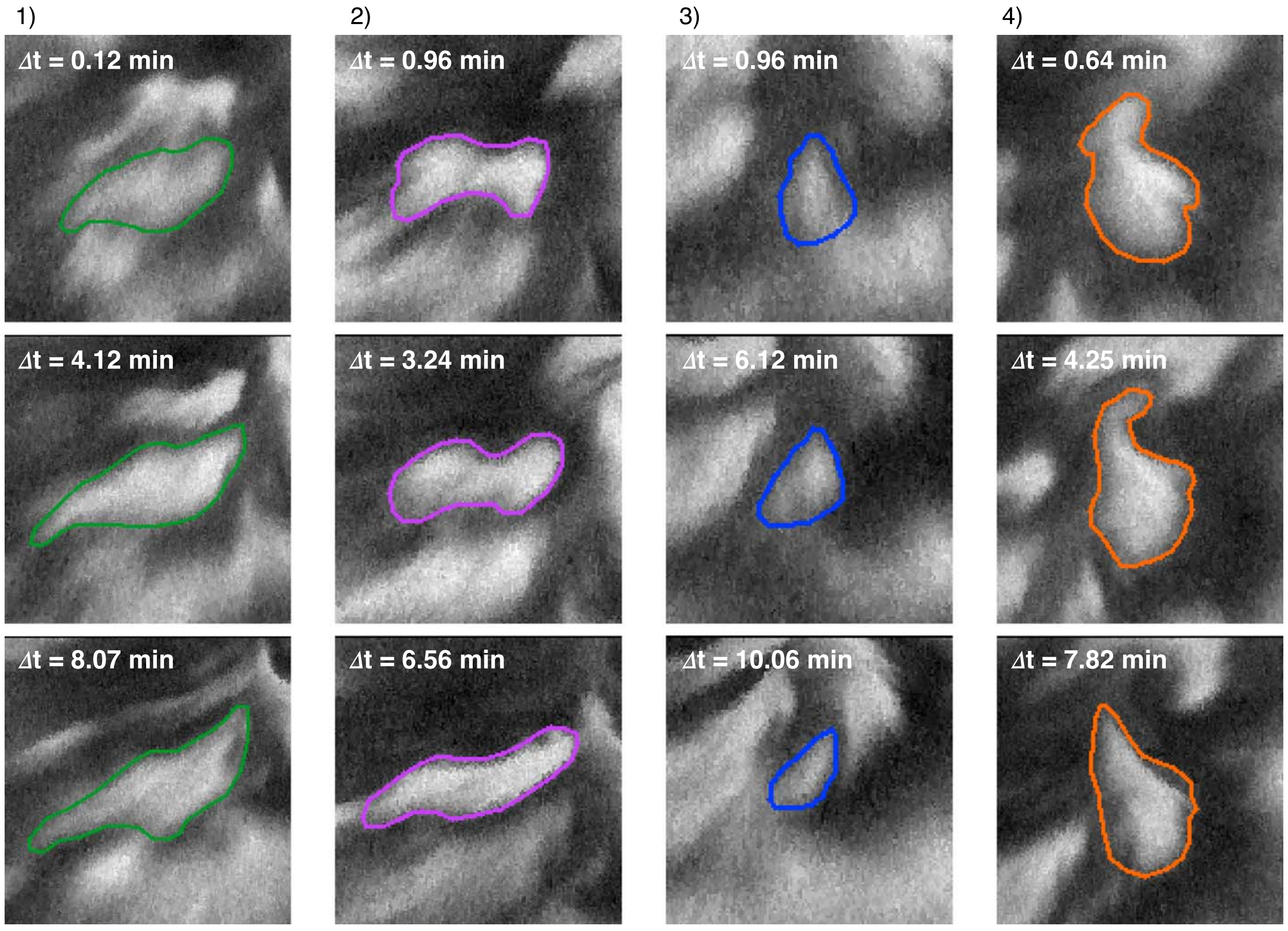

**Figure 3.** For each of the patches (1–4): Examples of contours that are manually identified by drawing a line around the patch (from top to bottom), where $\Delta t$ is the time from start of the analysis (relative to the time axis on Figure 2). The patch can then be extracted as all pixels within the closed contour.

### 3.2. Fixed Point Analysis

The advantage of studying fluctuating auroral patches from ground-based ASI is the ability to untangle the space-time ambiguity by removing the apparent velocities of the patches in the ASI frame of reference. The velocity of a patch in the ASI frame of reference is due to a combination of the ASI rotating with the Earth (moving the FOV across the night sky) and the drift of the patch in an inertial frame [effectively Geocentric Solar Ecliptic (GSE) coordinates]. To determine the apparent drift of the patch, we make an outline around the patch and find the velocity that best follows the patch through its lifetime. This technique assumes that the velocity is constant during the lifetime of the patch (~10 min), which none of the patches appear to violate. Using the apparent drift velocity, a new movie is made where the patch is centered at all times.

### 3.3. Contour and Extract the Patch

The complication of adjacent patches render automated techniques for patch boundary determination difficult. A manual determination therefore proved to be the more reliable technique. The extracted patch is expected to vary during the train of intensity fluctuations between off/dim and on/bright. A new contour of the patch is therefore determined at the peak intensity of each fluctuation as determined from the median of a square box within the patch. The contours are manually identified by drawing a line around the patch. In this way we can consistently avoid capturing adjacent patches. The brain is brilliant at identifying patterns, but exactly where the contour lies can of course be somewhat subjective. The contours are therefore validated by an independent control of where the contours are drawn. Figure 3 shows three example contours for each of the patches (1–4) from the beginning, center, and last part of the time interval analyzed. For each of the contours the time from start of the analysis ($\Delta t$) is given so that it easily can be compared to the position-time diagrams in Figure 2.

### 3.4. Quantitative Characteristics

Following the above selection, the patch can easily be extracted as all pixels within the closed contour. Each contour is considered valid over the fluctuation. If we are not able to draw a convincing contour, for example, because adjacent patches are on/bright, the previous and following contours are interpolated using a square function. In this way we can easily find the patch intensity defined as the total emissions within the patch (counts can be converted to kR using the relationship: $R = (\text{counts} - 350) \times 9.16$, where 350 is the imager dark counts).

The fluctuations are found from local minima after applying a boxcar filter and a fluctuation threshold (see Humberset et al., 2016, for a more thorough explanation).

The drift of the patch in an inertial frame (effectively GSE coordinates) is found by correcting the velocity of the patch in the ASI frame of reference for moving with Earth's rotation (see section 3.2). The velocity of Earth's rotation is easily calculated from the Earth's rotational period and distance at the geographic latitude and altitude of the patches.

The characteristics of the time variation and energy deposition of the patches are provided in an earlier study by Humberset et al. (2016). However, note that the more accurate manual contouring of this study allows us to extend the interval of analysis and to use the total emissions instead of the median emissions within the patch.

## 4. Results

To answer the stated science objective, we first give an overview of the patches before we show the persistence of patch shape and whether the entire patch fluctuates in a coherent fashion. At last we compare the patch drift to the $\vec{E} \times \vec{B}$ velocity.

### 4.1. Overview of Patches

In this study we analyze four patches with a considerable size ranging from ~1,000 to 5,000 km$^2$ (assuming an altitude of 110 km). All are relatively well defined with a sharp boundary to the dark sky or diffuse background emissions. However, the fluctuating auroral display is very complex with adjacent patches, possible spatiotemporal variations/streaming within the patch and relative movement between the patches. Even patches that at first seem like relatively well-separated individual patches are difficult to trace over a considerable part of their lifetime. We therefore start with an overview of each of the four patches.

#### 4.1.1. Patch 1

Patch 1 is a large patch (~4,800 km$^2$) that we analyze for about 8.5 min (15:19:03–15:27:30 UT). Patch 1 can be identified in the movie about 15 min earlier, but it is obscured by adjacent patches. The shape persists for some time after we end the analysis, but it is increasingly difficult to decide if Patch 1 has merged with adjacent patches or if it has split into several patches.

Figure 4 shows an overview of Patch 1 over the 8.5 min during which we can confidently contour and extract it as an individual patch. It displays eight example images, where Patch 1 is alternating between either on/bright or off/dim, and the train of fluctuations in intensity (total emissions within the patch). The dashed lines indicate the time and patch intensity of the example images. In the first example image (Figure 4a) the patch is on/bright and surrounded by several nearby and adjacent patches that also are on. The adjacent patches to the south (geographically) are on/bright, while the adjacent patches to the north-east and north are off/dim. In the second image (Figure 4b) the patch is off/dim, while the adjacent patches to the south remain, and the adjacent patch to the north-east is relatively bright. In the next images (Figures 4e and 4f), Patch 1 has moved away from the southern patches and it is easier to draw the contour. In the last example image (Figure 4h) the patch has entirely disappeared. Note that the patch typically is slightly brighter than the surrounding background even when it is off. This can be seen in the other example images (Figures 4b, 4d, and 4f), where especially the eastern part of the patch remains with slightly higher emission than the surrounding background. As can be seen from the train of intensity fluctuations, they are in general highly irregular both in duration and intensity. This led Humberset et al. (2016) to suggest that pulsating aurora is not an appropriate term and suggested "fluctuating aurora" instead. The patch shape may be the only parameter that remains relatively constant.

#### 4.1.2. Patch 2

Patch 2 (~2,600 km$^2$) is analyzed for 7.25 min (15:19:00–15:26:14 UT), as shown in Figure 5. It can be identified about 16 min earlier, but it starts out as two patches that are difficult to separate from adjacent patches. In the first example images (Figures 5a and 5c), the patch is surrounded by several adjacent patches, which makes it difficult to decide where to draw the contour. The adjacent patches to the south (geographically) are on/bright, while the adjacent patches to the north-east and north are off/dim. These adjacent patches disappear after the first minutes, while two new adjacent patches appear later in the interval. The first is south (bright in image e) of the patch, while the second is east (visible but relatively dim in image g) of the patch. These are, however, easy to avoid when drawing the contours.

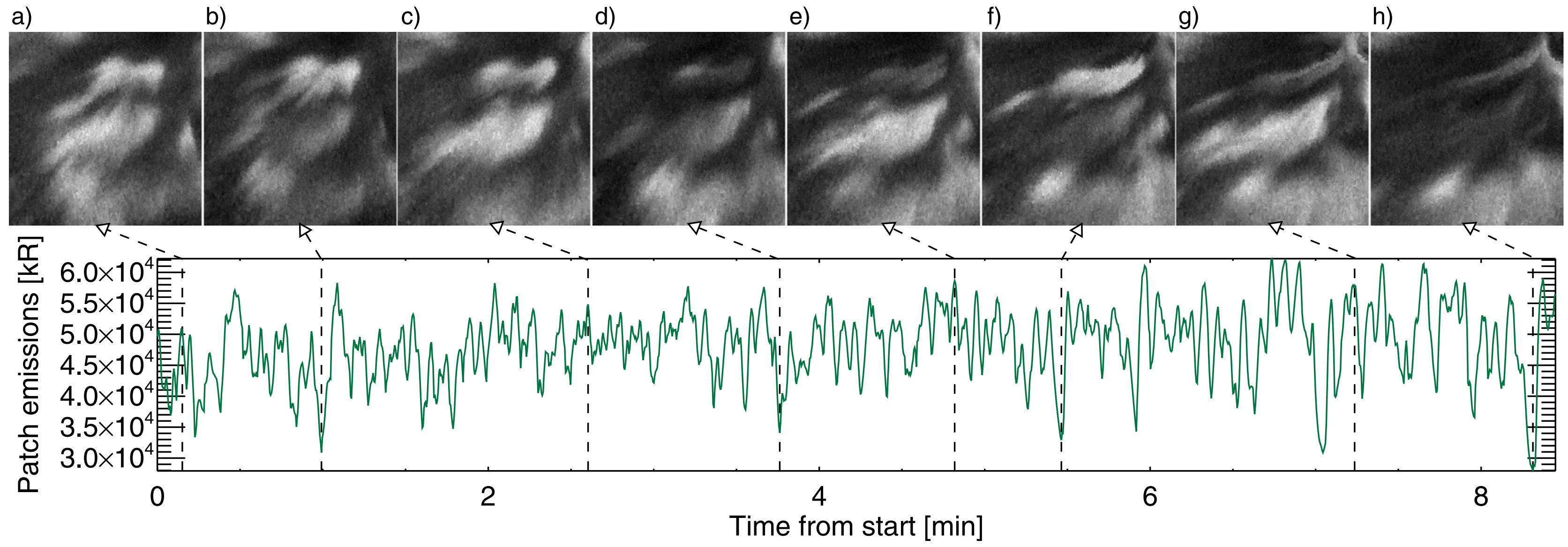

**Figure 4.** Overview of Patch 1. The train of fluctuations in intensity (total emissions within the patch) over the time interval that is analyzed and eight example images where the patch is either considered on/bright or off/dim. The dashed lines indicate the time and patch intensity of the example images.

Patch 2 also has some intriguing irregularities in the fluctuations across the patch, where one side of the patch is bright while the other side of the patch is dim/off. This can be seen around 1.8, 2.4 and 3.5 min on the time axis of panel 2 in Figure 2.

We stop the analysis when Patch 2 drifts close to Patch 4 (west/right in example image g). The patch emissions are in general highly erratic both in duration and intensity, while the patch shape stretches in the east-west direction and slightly rotates.

#### 4.1.3. Patch 3

Patch 3 is a relative small patch (~1,000 km$^2$) that we follow for about 10 min (15:17:05–15:27:12 UT). The patch emission and example images, where the patch is considered either bright/on or dim/off, are shown in Figure 6. At first glance it seems to be related to the smeared out southern patches (south in the example images). However, a visual inspection of the movie and several position-time diagrams leaves no doubt that Patch 3 is an individual patch from about 15:17 UT. We choose to start the analysis when it is still very close to the southern patches (example images a–c), which of course makes the drawing of the first contours a little uncertain. The relative movement then causes the patch to split from the southern patches and get closer to the other nearby patches 1 (to the east) and 4 (to the northwest). Throughout the interval the patch emissions show a complex fluctuating behavior, while the patch shape appears to slowly rotate clockwise. Toward the end of the interval the patch emissions are dimming. The patch occurs a few more times after we stop the analysis but the fluctuations are very dim.

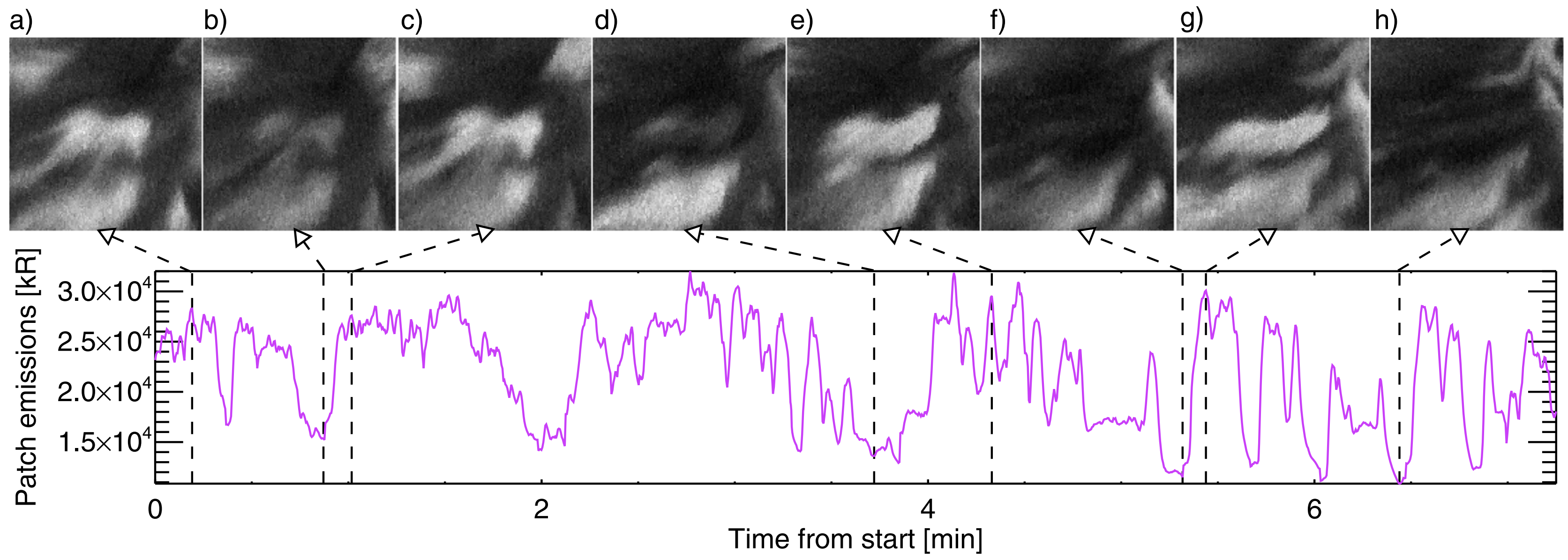

**Figure 5.** Overview of Patch 2 displaying the same parameters as Figure 4.

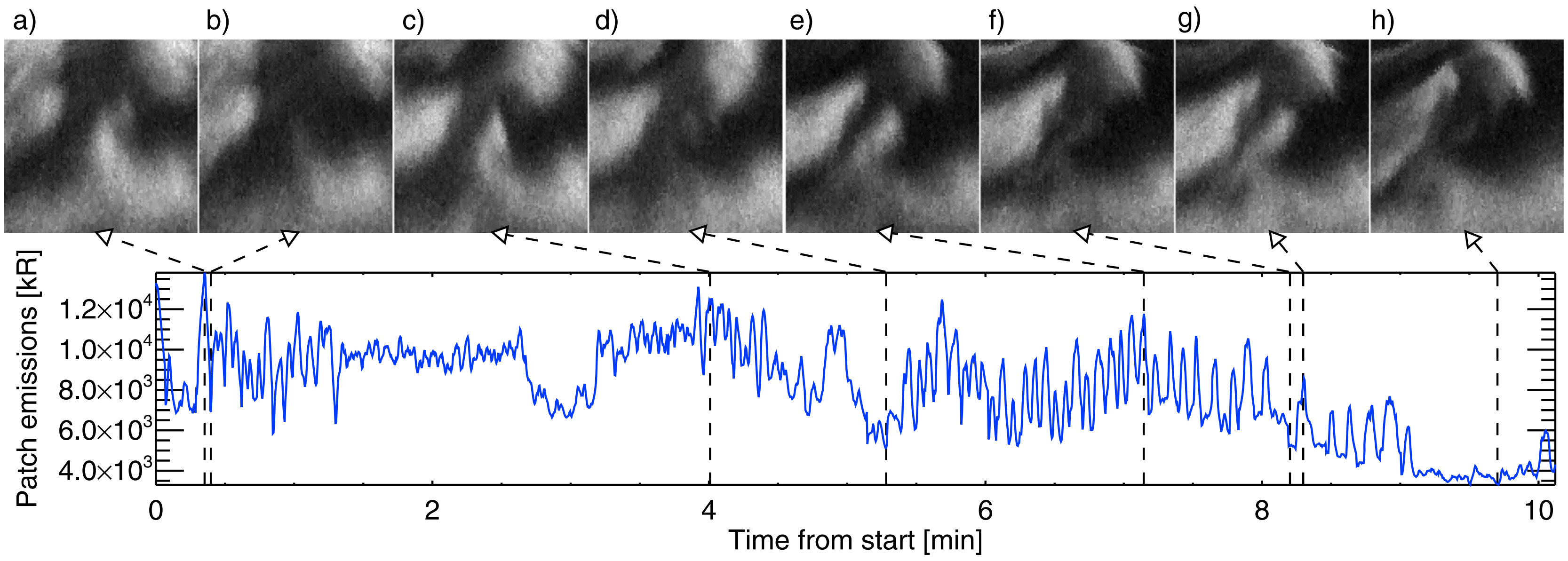

**Figure 6.** Overview of Patch 3 displaying the same parameters as Figure 4.

### 4.1.4. Patch 4

Patch 4 (~3,300 km$^2$) starts out as a complex patch system. We start the analysis when it has become well defined and follow it for the ~9 min (15:17:14–15:26:02 UT) that is displayed in Figure 7.

Patch 4 is the toughest of the patches to contour, mainly due to two features. The first feature is the smeared-out eastern edge of the patch where the emissions falls off gradually. This makes it difficult to decide where the contour should be drawn, especially in the end of the interval when the nearby patches have come closer (see example image g in Figure 7). The second feature is the northern part of the patch, which has a difficult shape to contour and where an adjacent patch fills the gap between Patch 4 and the nearby patch to the north (see image a). At 6.0–6.7 min into the time interval, the northern adjacent patch fluctuates in the same manner as Patch 4 (see image e) before it fades away. However, when it brightens again about 30 s later, it has joined together with the northern part of Patch 4 and starts to fluctuate in a different manner to the main part of Patch 4 (see images g and h). We continue to follow Patch 4, where we leave out the northern part which has separated. This is evident in the drop in the patch emissions in Figure 7.

Patch 4 has some intriguing irregularities in the fluctuations across the patch, and a core which fluctuates less than the rest of the patch, as can be seen in the position-time diagram of Patch 4 (Figure 2). There is however no obvious pattern. The patch can vary or retract and stream out (southeastward and northward) several times before the whole patch is dimming. When it brightens, the core sometimes turns on first, while at other times the southeastward patch brightens first. Another interesting irregularity appears about 5.5 min into the time interval, where the edges of the patch (east-west) continue to fluctuate but are very dim compared to the bright core. This can be seen in panel 4 of Figure 3, where the contour at $\Delta T$ = 7.82 min from start looks like it is drawn too wide because we include the dim edges.

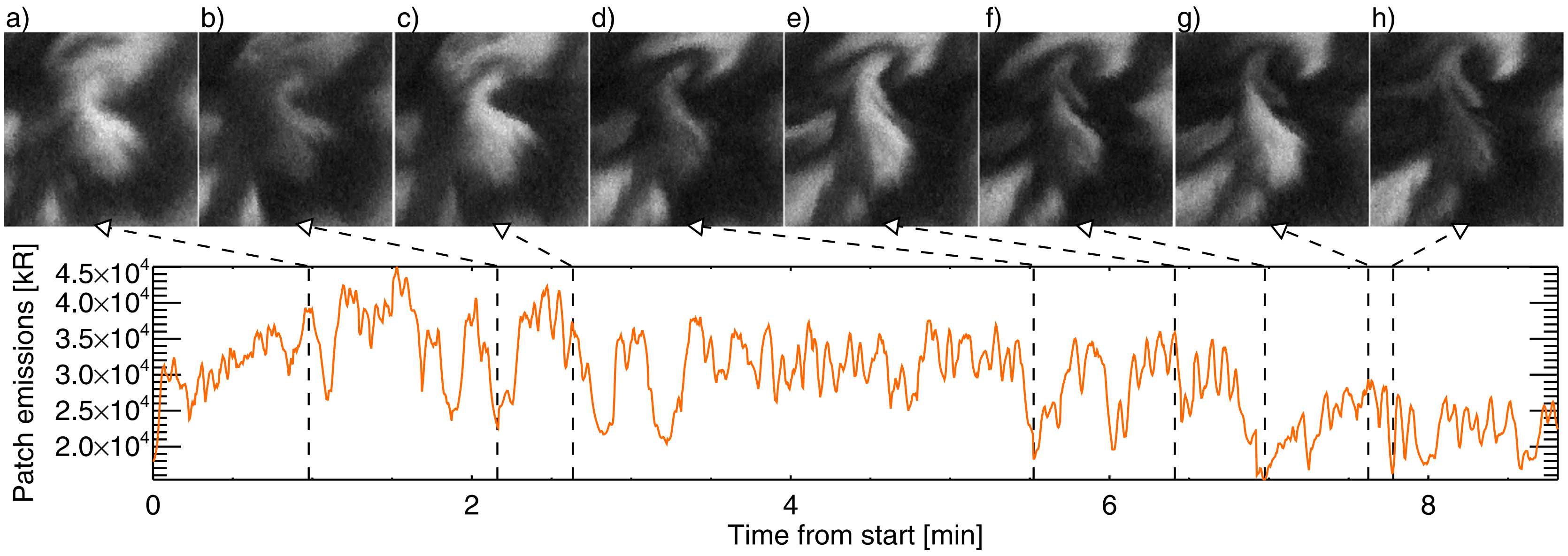

**Figure 7.** Overview of Patch 4 displaying the same parameters as Figure 4.

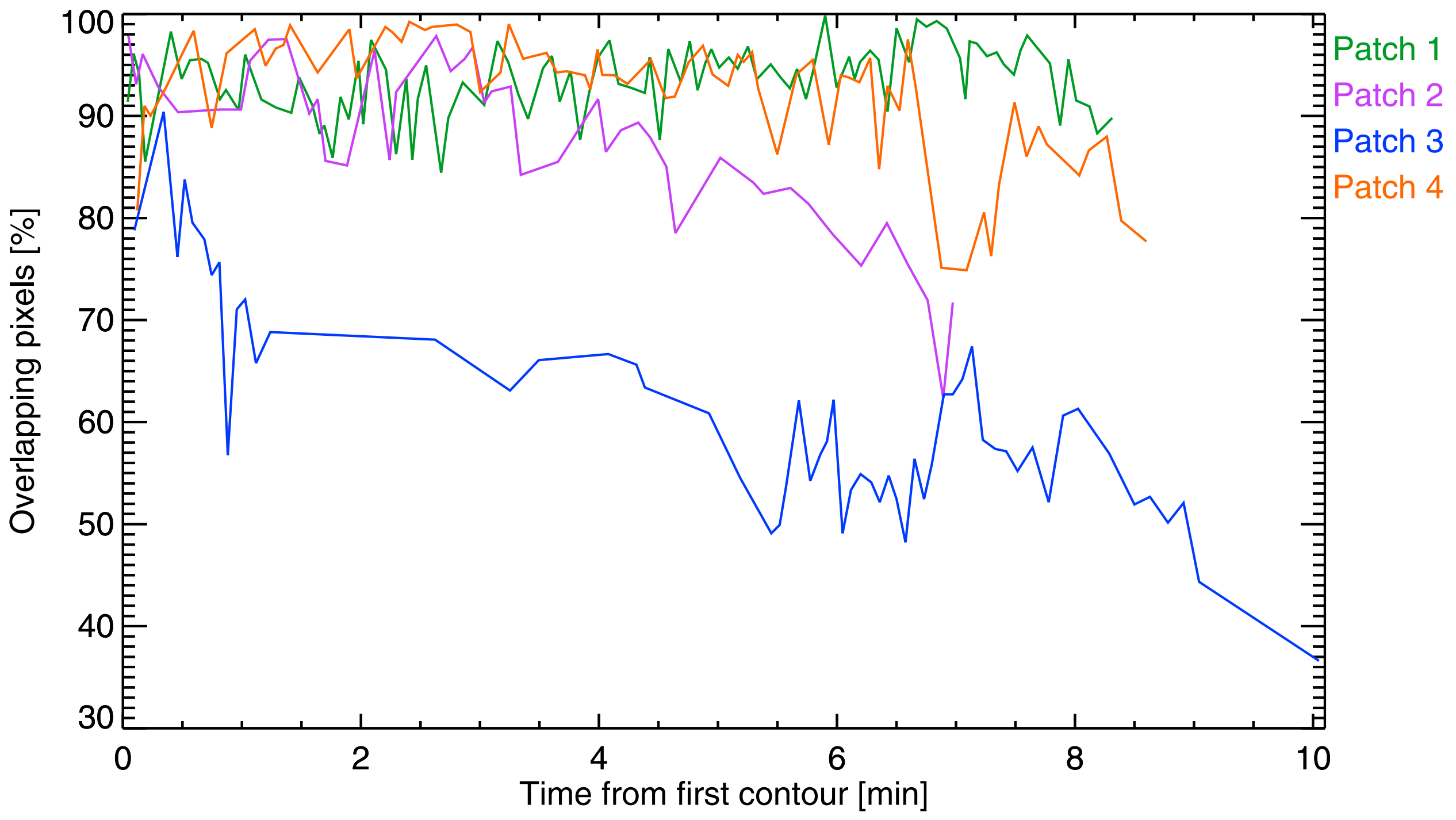

**Figure 8.** Percentage of patch (pixels within the contour) that overlaps with the patch at the first fluctuation.

Similar to the other patches, also the Patch 4 emissions are in general highly complex both in duration and intensity. It is apparent that during the off-time, the patch still has emissions slightly higher than the surrounding background. Except for the detaching of the northern part, the patch shape remains relatively constant.

### 4.2. Persistence of Patch Shape

To what extent does a patch maintain its shape? The fluctuating auroral patches are often described as persistent. Exactly how persistent is often loosely described, and there are no quantitative measures to how the shape of a patch evolves. We therefore provide a quantitative and qualitative estimate of the persistency of the patch shape.

Figure 8 gives a quantitative estimate of how persistent the patch shape is. It shows the temporal evolution of the shape in the patch frame of reference as the percentage of the patch (pixels within the contour) that overlaps with the patch at the first fluctuation of the analysis interval. The shape of Patch 1 is remarkably constant (85–100%) throughout the analysis interval of about 8.5 min. Patch 2 gradually drops to about 70% overlap. Patch 3 drops to 60–70% within a minute, and then the overlap drops gradually for the next 9 min until it reach as low as 40% overlap. Patch 4 also has a remarkably constant shape (90–80% until about 7 min when part of the patch detach and there is a small drop in the overlap.

Figure 9 vividly illustrates how remarkably persistent the shape of each patch (1–4) is. It shows 20 of the contours, including the first and last contours, in the patch frame of reference. The temporal color scale visualizes the evolution of the patch shape. The shape of Patch 1 stretches in the magnetic east-west direction, but remains remarkably similar over the ~8.5 min we follow it. Patch 2 also stretches in the geographic east-west direction and rotates slightly from the geographic to the magnetic east-west direction (counterclockwise). Patch 3 rotates clockwise from geographic north-south and shrinks in size the last minute before it fades away. The shape of Patch 4 evolves into a more narrow shape and the curl of the northern part stretches until it detaches for the last ~1.5 min.

### 4.3. Fluctuations Within the Patch

Does the entire patch fluctuate in a coherent fashion? If the answer is yes, that would imply that all pixels within the patch are highly correlated with each other and, specifically, that this correlation is independent on distance between the pixels. Figure 10 shows the results from a straightforward analysis aimed at answering this simple question. We pick a number of pixels along a line that is within the patch at all times. For each pixel we determine the time series and calculate the correlation coefficients between pixel at the beginning of the arrow and the other pixels along the colored line. Figure 10 shows the average of the correlation coefficients for all fluctuations in the analysis interval as a function of the distance between the pixels.

As the figure clearly shows, the correlation coefficient has a pronounced dependence on the spatial distance, except for the relatively small Patch 3. If we consider coherency between fluctuations to have a correlation coefficient above 0.5, Patch 3 fluctuates in a coherent fashion, while the fluctuations of patches 1 and 4 could be considered coherent for about 15 km and Patch 2 for about 50 km. However, regardless of correlation value the patches 1, 2, and 4 appear to indicate a nearly linear dependence on distance. Thus, for only one of four patches do we find that the entire patch fluctuates in a coherent fashion.

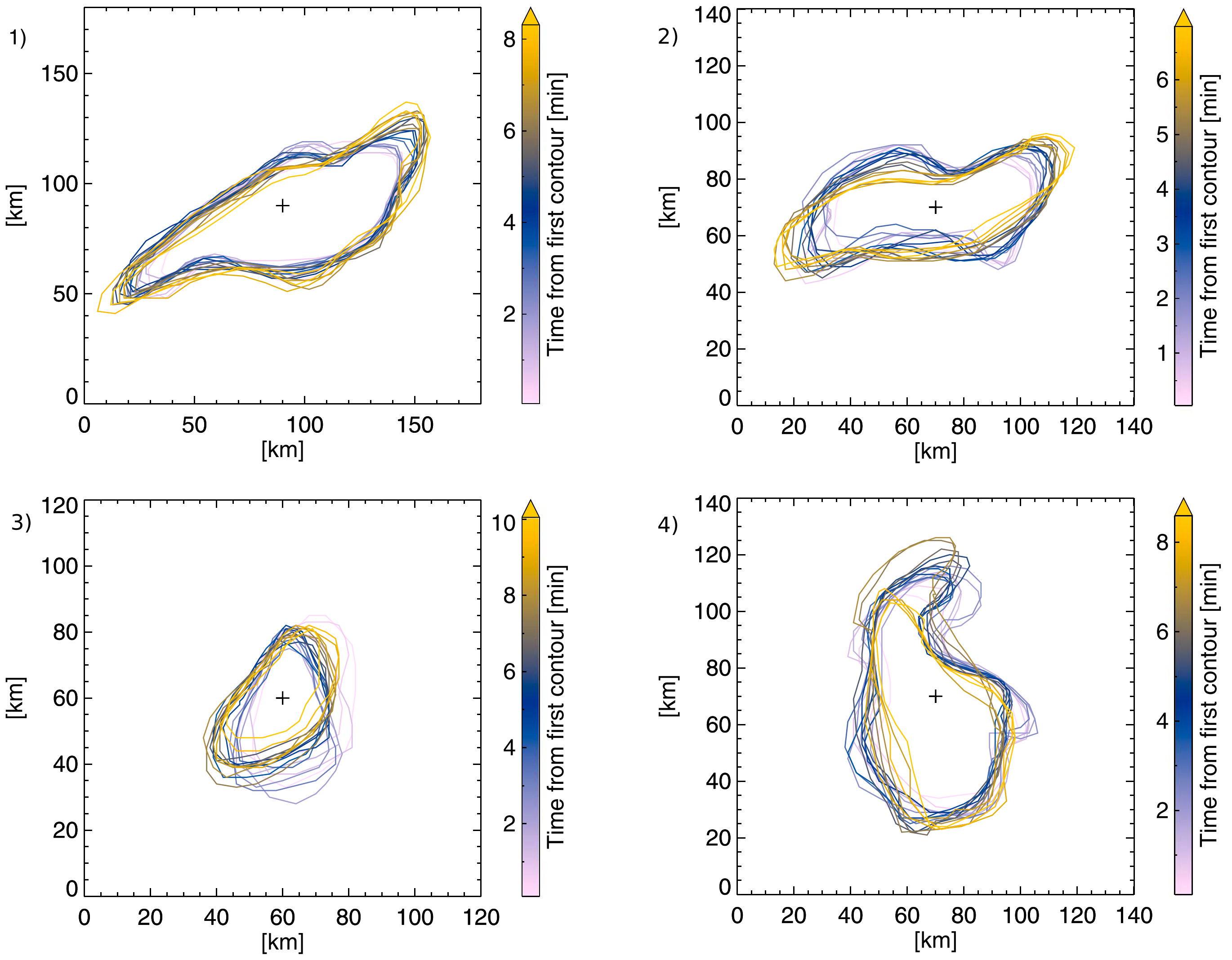

**Figure 9.** For each of the patches (1–4): Twenty of the contours, including the first and last contours, in the patch frame of reference. The temporal color scale visualizes the evolution of the patch shape.

### 4.4. Patch Drift

It has been suggested that the fluctuating auroral patches are drifting with the ionospheric $\vec{E} \times \vec{B}$ drift (e.g., Scourfield et al., 1983; Yang et al., 2015, 2017). The velocities of the patches are found in the geographical oriented Cartesian grid and rotated to the Altitude Adjusted Corrected Geomagnetic (AACGM) coordinate system. The drifts of the patches in a reference frame not corotating with the Earth are in the range of 230–287 m/s in a north-eastward direction.

As an estimate of the $\vec{E} \times \vec{B}$ velocity we use the SuperDARN (Greenwald et al., 1995) data provided by Mr. Robin Barnes at the Johns Hopkins University Applied Physics Laboratory using the technique described by Ruohoniemi and Baker (1998). The resulting $\vec{E} \times \vec{B}$ convection velocities, also called fitted velocities, are found in a reference frame corotating with Earth. The $\vec{E} \times \vec{B}$ velocities are therefore compared to the patch drift velocities that are not corrected for the ASI rotating with Earth.

In Figure 11 we show the patch velocities and the median of the $\vec{E} \times \vec{B}$ (SuperDARN) velocities at the location of each of the patches. We find that all patches move in a north and slightly eastward direction with a speed of 53–104 m/s. According to the SuperDARN results the convection is mostly eastward with a speed of about 140 m/s. The direction of the patches and the ionospheric plasma differs by 51° to 73°. Assuming that the SuperDARN drifts are correct, this result indicates that the fluctuating patches drift independently from the ionospheric plasma.

## 5. Discussion

We first address the inherent limitations of the ASI data set and technique, and then we discuss to what extent the fluctuating auroral patches maintain their shape and fluctuate in a coherent fashion, and if the patches always drift with the $\vec{E} \times \vec{B}$ velocity.

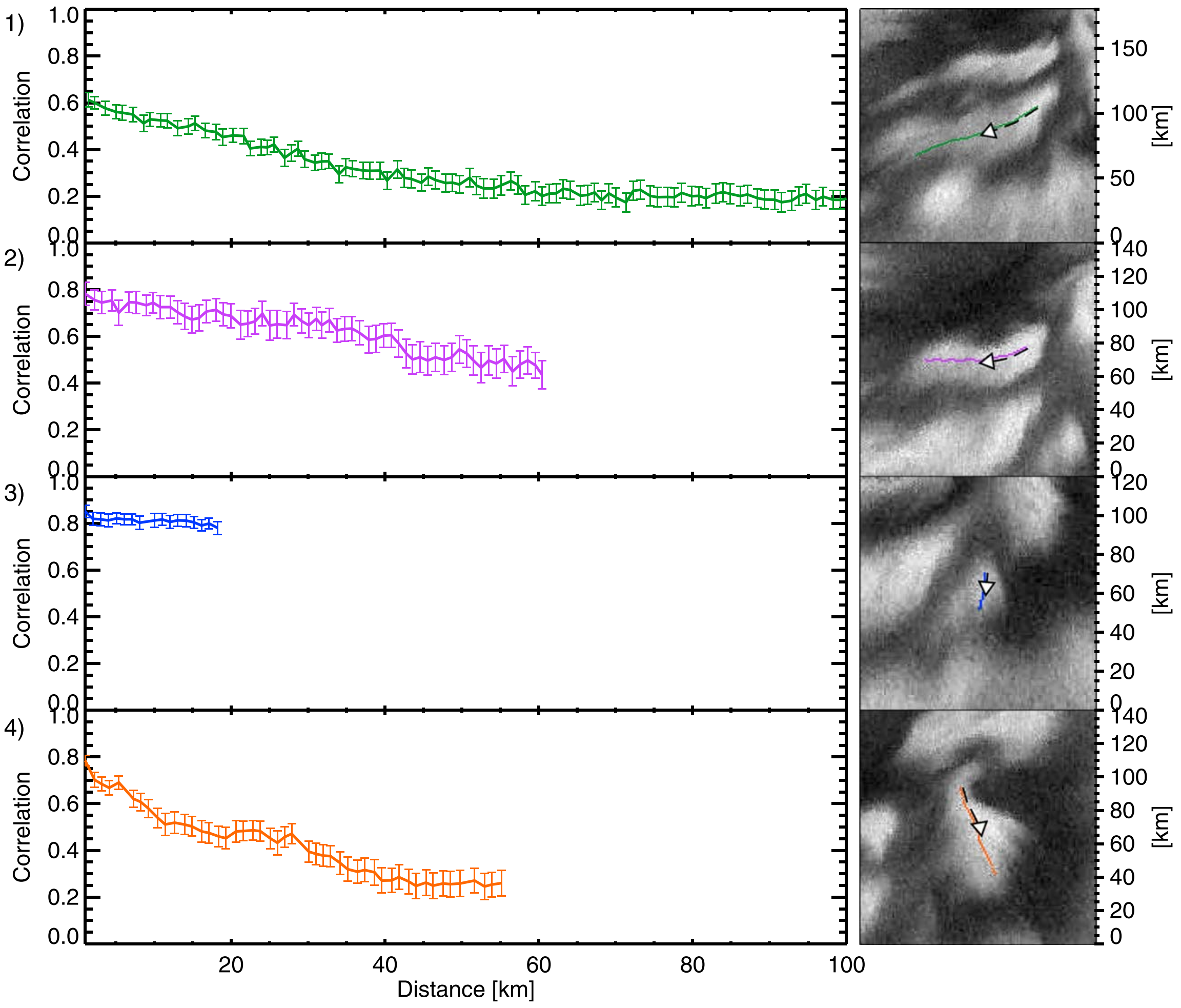

**Figure 10.** For each of the patches (1–4): The average correlation between all fluctuations of the first pixel to the next pixels as a function of the distance between them. The standard error is shown as error bars. The pixels are sampled along the line which is highlighted in color on the example image to the right. The arrow indicates the increasing distance and thus the first pixel.

## 5.1. Inherent Limitations

The main limitations are as follows: (1) observations are limited to four patches, and the analysis is limited by (2) the camera capabilities, and (3) the fact that we only study the 557.7 nm emissions.

### 5.1.1. Limited Number of Patches

The fluctuating auroras are observed under a wide variety of conditions (e.g., Jones et al., 2011), while this study is limited to effectively one set of conditions. Of most importance, the patches are located in the dawn sector at ~4 MLT during substorm activity. The patchy type of fluctuating auroras are most frequently observed toward dawn (Cresswell, 1972). The limited observations of course result in limited statistics, and we cannot address how the characteristics depend on local time, geomagnetic conditions, and so on. At the same time this means that any variations between the four patches are not due to such dependencies.

### 5.1.2. Camera Capabilities

The camera is an all-sky imager sampling images at a time rate of 3.31 Hz. An all-sky imager has a large FOV providing good coverage of the auroral display, but it has some considerable distortions. The main issue is that ASI obtains the column integration of auroral emissions that originate from a range of altitudes and magnetic field lines. Additionally, there are distortions due to optics and elevation (look direction). The question is whether smearing affects our ability to answer the stated science objective.

The common height assumption of 110 km of the 557.7 nm aurora does not have any considerable effects on the Earth's rotation velocity or on the ASI pixel size, which is much more affected by the elevation angle. There is no robust way to remove smearing. Therefore, we mediate the smearing by evaluating patches within the center 400 by 400 km FOV. The resampled image has a pixel size of 1.0 km. Based on the size of the largest pixels in the fish-eye view that are projected onto the center FOV of the resampled grid, we determine that the smearing is less than ~2.5 km.

The 3.31-Hz time resolution of the ASI data is limited by the Nyqvist frequency of 1.65 Hz. We can therefore not distinguish rapid temporal behavior such as ~3 Hz modulations often found to be superimposed on the slower on-off fluctuations. Running the all-sky camera at a higher cadence is possible, but the fluctuating aurora that occurs with higher-frequency fluctuations, such as the ~3 Hz modulations, likely occurs at small spatial scales on top of the large-scale fluctuations (Nishiyama et al., 2016). A narrow field of view imager is therefore more appropriate for investigating the spatiotemporal variations of the higher frequencies, as was done by Samara and Michell (2010), which reported fluctuations up to 10–15 Hz. It is possible that these higher-frequency fluctuations are caused by completely different mechanisms, and therefore, it would make sense to investigate them independently.

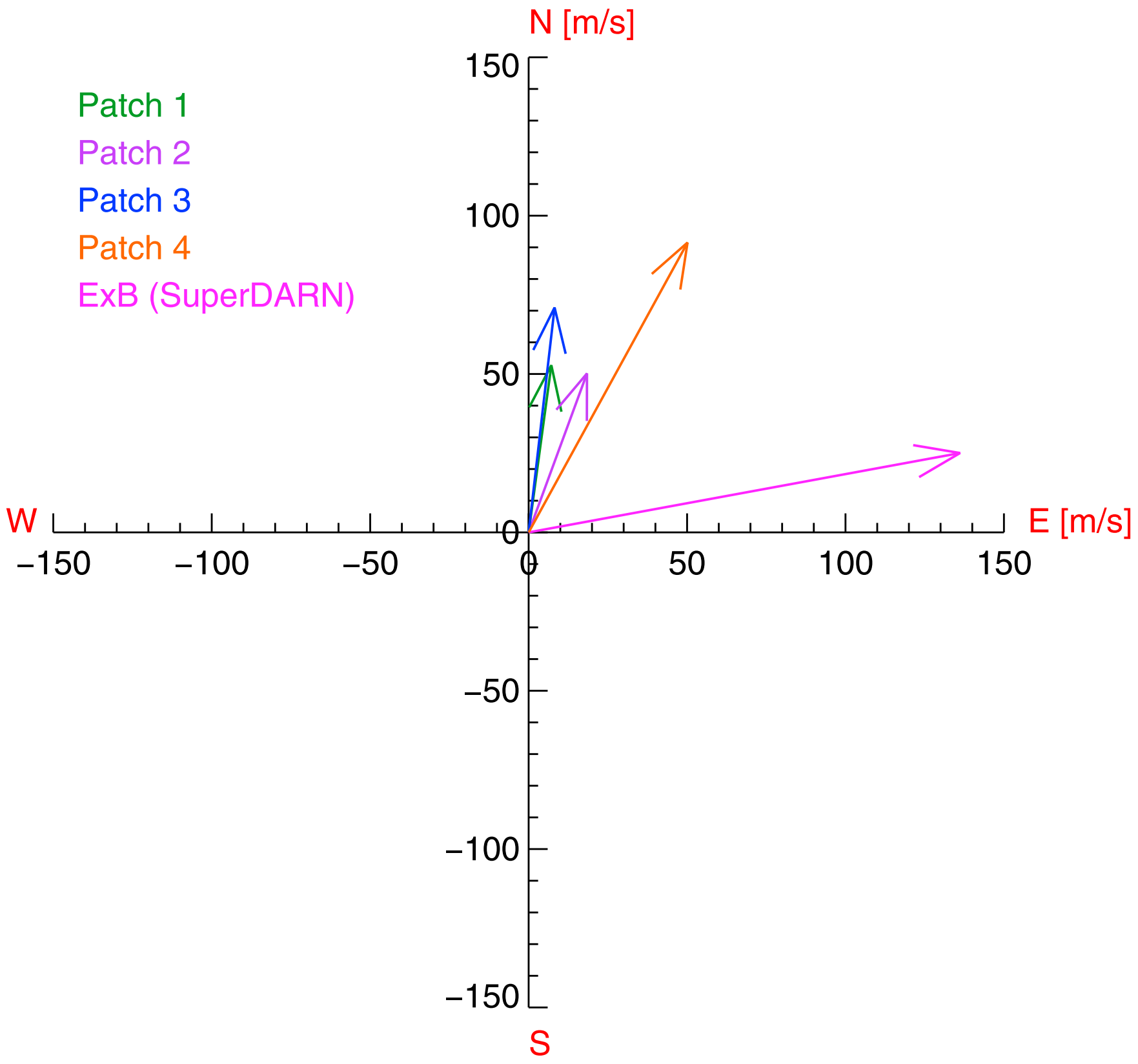

**Figure 11.** The drift velocities of the patches compared to the $\vec{E} \times \vec{B}$ velocity (SuperDARN) shown in the AACGM magnetic coordinate system (corotating). SuperDARN = Super Dual Auroral Radar Network; AACGM = Altitude Adjusted Corrected Geomagnetic.

#### 5.1.3. The 557.7 nm Emissions

The 557.7 nm emissions are limited by chemical effects. Its green-light is the brightest line in the visible spectrum in most auroras, and also for fluctuating aurora. It is however limited by a mean effective lifetime of 0.3 to 0.59 s of the O($^1$S) excitation, the highest values found in observations of sharp-edged fluctuating patches (Scourfield et al., 1971). This is due to the ~0.75 s lifetime of the direct O($^1$S) excitation, quenching of the excited state below 100 km and additional indirect excitation processes (e.g., Brekke, 2013). The result is a time lag to the prompt emissions (e.g., the blue 427.8 nm) and a temporal smoothing over the same time scale, meaning that the 557.7 nm filter used could smooth out possible impulsive behavior. Also the O($^1$S) excitations resulting from the highest-energy electrons (>20 keV) that penetrates to altitudes below 100 km could be quenched and therefore become less visible in the 557.7 nm emissions. We are however investigating sharp-edged fluctuating patches. So if the observations by Scourfield et al. (1971) can be generalized, the high values of mean effective lifetime of the O($^1$S) excitation can indicate that we capture most of the fluctuating precipitation.

In addition, impulsive behavior from a source region far from Earth can be smoothed by the energy-time dispersion due to the energy-dependent transport times along the field line connecting the magnetosphere and the ionosphere. This was discussed by Humberset et al. (2016) that found the most probable time lag in the beam of energetic electron precipitation to be 0.6 s or less. It is therefore likely that the spatiotemporal characteristics we observe are due to the behavior of the underlying source mechanism and not a dispersion effect.

### 5.2. To What Extent Does the Patch Maintain Its Shape?

The fluctuating auroral patches are often described as persistent (e.g., Johnstone, 1978; Oguti, 1976). We found a quantitive measure (Figure 8) to the persistency of the patch shape in the form of percentage overlapping pixels. Patches 1, 2, and 4 all have a remarkably persistent shape with above 85% overlap for 4.5–8.5 min. The exception is Patch 3 for which the overlap drops to ~60% in about a minute. However, Patch 3 might go through a transition and then remain relatively constant throughout the rest of the interval. In Figure 12 we test this by showing the percentage overlap compared to the preceding patch shape. The percentage of overlapping pixels no longer drops to a low value but instead appears to vary between 85 and 100%. This means that Patch 3 indeed goes through a transition in the first minutes, and a closer inspection of the contours in Figure 9 reveals that the transition is due to a change to the northern part of the patch (the before relatively dim northwestern part fades away). All of the patches can therefore be considered to have remarkably persistent shapes with above 85% overlap over periods ranging from 4.5 to 8.5 min.

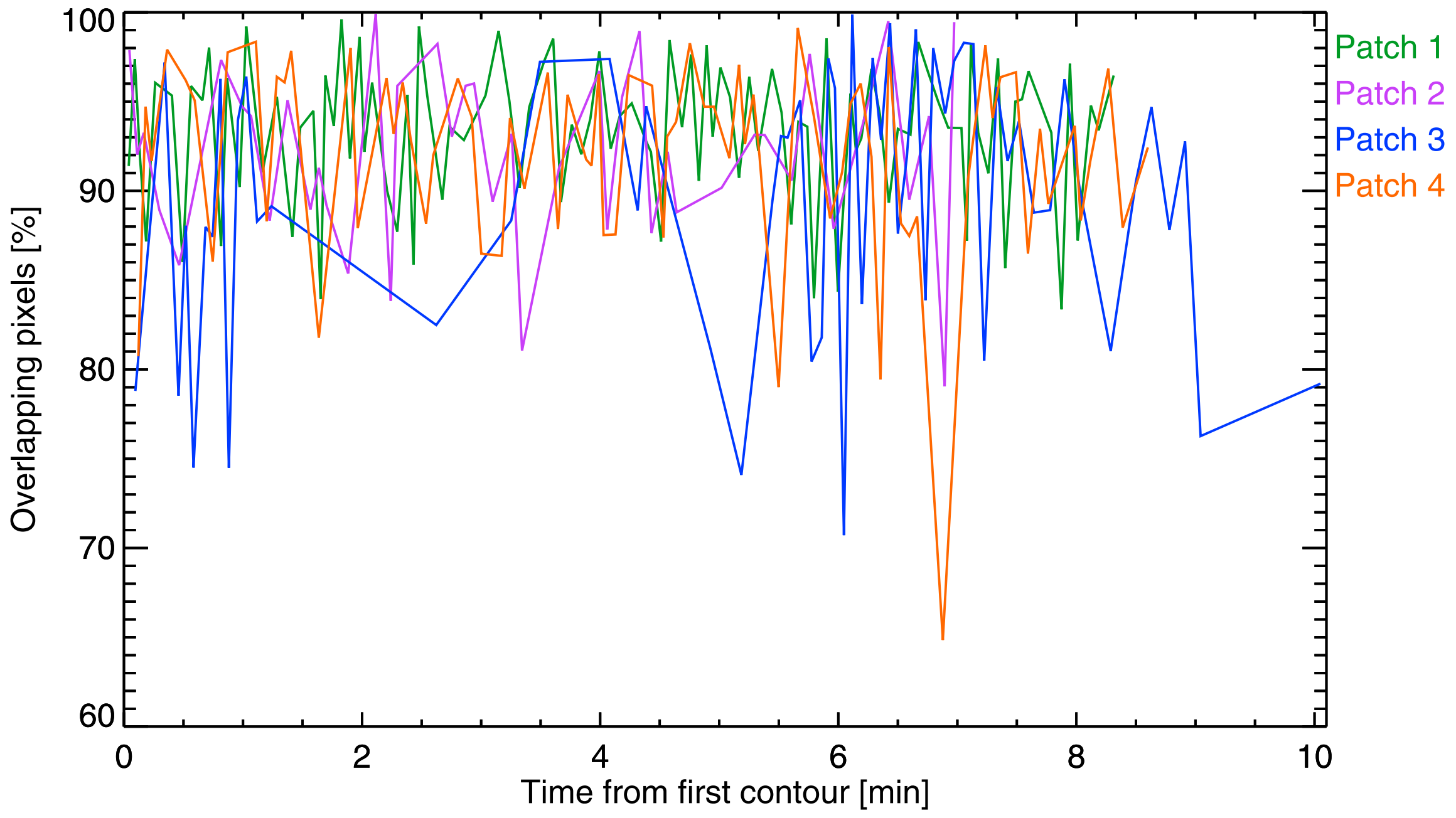

**Figure 12.** For each of the patches (1–4): Percentage of patch (pixels within the contour) that overlaps with the patch at the peak of the latter fluctuation.

#### 5.2.1. Qualitative Evolution of Patch Shape

In our analysis we have not accounted for any rotation of the patch. If we adjusted patches 2 and 4 for rotation, their shapes could be considered as persistent over a longer time than what is evident from Figure 8. As can be seen in Figure 9, the patches mostly undergo gradual changes and largely maintain their shapes over the time we follow them. Either they get slightly more elongated and narrow, rotate, or a combination of those. We also note that three of the patches (1–3) rotate and elongate toward the magnetic east-west direction, while one of the patches (4) does not seem to obey the same trend.

#### 5.2.2. Implications

This is the first study with a quantitative description of patch shape evolution. It is suggested that the shape of fluctuating patches is governed by the wave resonance/cold plasma region at the magnetospheric equator, and that cold plasma of ionospheric origin acts to keep the region stable (e.g., Li et al., 2012; Liang et al., 2015; Oguti, 1976). Alternatively, it is suggested that conductivity gradients in the ionosphere due to the energetic electron precipitation can modify the shapes of fluctuating auroral patches (Hosokawa et al., 2010). There are however no detailed predictions of patch evolution from these processes.

A remarkably stable patch shape implies two strict requirements: (1) The region of the underlying source mechanism must be stable, and (2) the mapping of the magnetic field lines connecting the patch to the magnetosphere must also be stable. With respect to the latter, this is baffling since these patches occurred during substorm conditions when the magnetosphere undergoes a large-scale reconfiguration. We speculate that the mechanism responsible for the visual part of the fluctuating patch is located not in the distant magnetosphere but much closer to the Earth (Sato et al., 2004), and thus is insensitive to changes in the magnetospheric field line morphology.

### 5.3. How Often Do the Patches Merge or Detach?

It happens that patches merge with one or several nearby patches. Exactly when and why this happens is not known, and neither is it clear what this implies for the underlying cause of the patch. We therefore give a summary of the merging and detaching that the above patches underwent.

Patch 2 starts out as at least two patches. It started to gradually merge into one individual patch from about 15:14 UT (5 min before we start the analysis). However, as described above, the two parts fluctuate in a slightly irregular manner 1.5–3.5 min into the time interval. Patch 2 moves relatively closer to Patch 4 and they merge at about 15:27 UT, a few tens of seconds after we stop the analyses. Also, about 2.5 min earlier the northern part of Patch 4 merges with an adjacent patch and detaches to form a new individual patch. We have not registered any merging and detaching of patches 1 and 3. In only 13 min two of the four patches participate in as much as three events of merging and detaching.

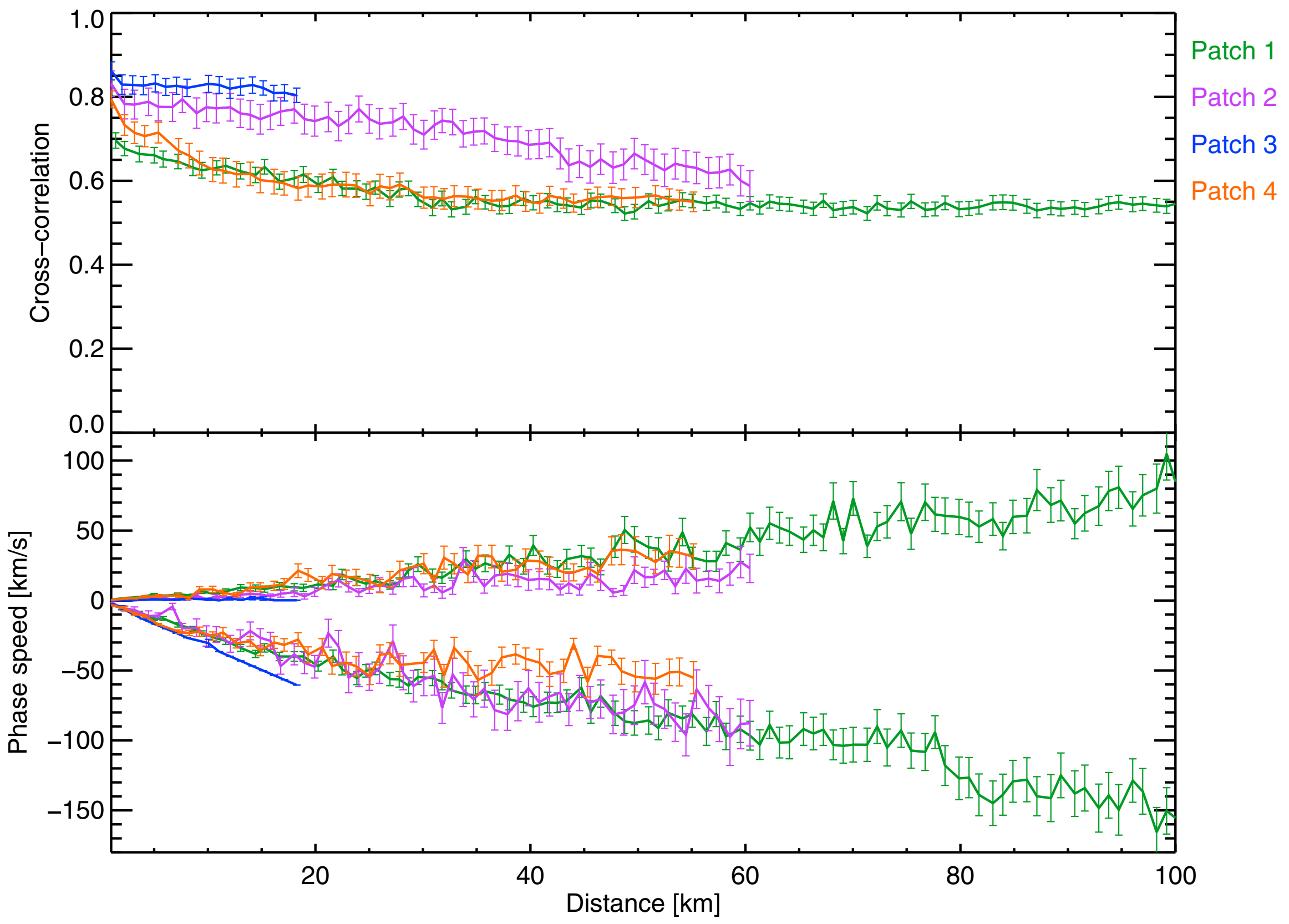

**Figure 13.** For each of the patches (1–4): The average of the best fit cross correlations and phase speeds between all fluctuations of the first pixel to the pixels sampled along the patch as shown in Figure 10. To avoid the possibility that positive and negative phase speeds can cancel each other out, the average of the positive phase speeds are found separately from the average of the negative phase speeds. The standard error is shown as error bars.

### 5.4. Does the Patch Fluctuate in a Coherent Fashion?

It is known that fluctuating aurora can occur in different spatiotemporal modes (see section 1.3). However, it is not known why the shapes sometimes fluctuate coherently and sometimes not, and the underlying cause of the spatiotemporal variations remains to be determined. We therefore ask whether the patches fluctuate in a coherent fashion. As seen in Figure 10, the three largest patches have temporal correlation coefficients that show negative dependencies on distance. Thus, only one of four patches fluctuates in a coherent fashion.

#### 5.4.1. Time-Delayed Response

As evident from Figure 10, the patches are clearly spatially variable, meaning that the intensity in one part of the patch might have higher/lower intensity and fluctuate less or stronger. To further investigate whether the spatial variation we see is due to so-called expanding or streaming mode fluctuating aurora, where the luminosity moves inside the patch due to a time-delayed response, we find the cross correlation and calculate the phase speed from the distance between the pixels and the time lag of the best fit of their fluctuations. To avoid the possibility that positive and negative phase speeds can cancel each other out, the averages are found separately for the positive and negative phase speeds. The results are shown in Figure 13. The correlation shows the same trend of negative dependence on distance as in Figure 10, but it is much less pronounced as the lowest correlation values have increased from 0.2–0.5 to 0.55–0.6. Thus, by shifting the time series we find a better fit. This is reflected in the phase speeds along the patch reaching as high as −150 km/s to 80 km/s for Patch 1, −90 km/s to 20 km/s for Patch 2, −50 km/s to 30 km/s for Patch 4. Even for Patch 3, which initially is considered relatively coherent, we find a better fit of the fluctuations by shifting them, resulting in negative phase speed of up to 50 km/s.

The trend of increasing phase speed per distance is likely a combination of two effects: (1) For small distances the speed is underestimated by our 0.3 s temporal resolution. (2) From the cross correlation we expect the fluctuations to be different, meaning that the best fit can result in a small shift that is not necessarily connected to a time-delayed response, but results in a relatively large phase speed. This can result in an overestimation of the average phase speed at large distances. The phase speeds we find are in agreement with earlier measured expansion speeds of ∼25–250 km/s usually uniformly in all directions, sometimes in a preferred direction (Kosch & Scourfield, 1992; Scourfield & Parsons, 1971). It is clear that the luminosity moves inside the patch due to a time-delayed response and that the patches possibly can be defined as streaming/expanding mode fluctuating aurora.

#### 5.4.2. Slower Varying Background Emissions

The fluctuations of the patches are superposed onto a background of slower varying emissions. This supports the description of Cresswell (1972) that the patchy type of fluctuating auroras generally undergo incomplete intensity fluctuations. The background of slower varying emissions than the intensity fluctuations are therefore treated as an offset and removed. The offset is found from the off/dim values to visualize the spatiotemporal variation of each fluctuation (build up and decay). Movies of the patches corrected for the offset (see the supporting information and Movies S2–S5) show the resulting spatiotemporal variation within the patch. We repeated the cross-correlation analysis (not shown) and do not find any significant changes from the results in Figure 13, and conclude that the background can be treated as mere offset. However, we notice that the offset corrected for the background emissions in the patch surroundings are of comparable brightness to the fluctuation. For example, for Patch 1 the median patch emissions is about 2.7 kR and the median offset is 2.5 kR, while Patch 2 has median patch emissions of about 3.8 kR and a median offset of 1.8 kR. Together with the observation that the background is not entirely removed between the fluctuations, this might support the suggestion that the background is not merely diffuse emissions but part of the mechanism (Dahlgren et al., 2017).

#### 5.4.3. Implications

It is not clear what causes the variation in fluctuation within the patch. The expanding/streaming spatial fluctuations are, for example, suggested to be accounted for by atmospheric waves (Luhmann, 1979) or by the pitch angle scattering moving to adjacent regions with lower densities of cold plasma as the energy of the waves grows (Tagirov et al., 1999). However, the auroral observational consequences are not quantified. The more recent studies are linking spatial structuring of fluctuating aurora to the higher-frequency fluctuations (>3 Hz) commonly found superimposed on the main emission fluctuation (Nishiyama et al., 2012; 2016; Samara & Michell, 2010; ). For example, Samara and Michell (2010) found that higher temporal frequencies exist when there are smaller-scale structures present, and Nishiyama et al. (2016) found that the rapid fluctuations were highly localized in substructures of the main form (propagating/moving mode). The spatial structuring was for various reasons found in agreement with chorus rising tone elements, but it is not clear if the rapid fluctuations are caused by completely different mechanisms than those causing the main on-off fluctuation. Neither is it clear whether the rapid fluctuations can explain the observed variation within the patch.

A closer inspection of the patch movies (see the supporting information and Movies S2–S5) show that the spatiotemporal variation is highly irregular. From one fluctuation to the next it varies what part of the patch turns on, or turns on first and in which order, and if there is expansion and/or subtraction and/or streaming. During one on-off fluctuation there can be several smaller fluctuations within a smaller area of the patch (sometimes the whole patch) that does not have sufficient a build-up or decay to be considered an individual fluctuation. We also find individual fluctuations that comprise only part of the patch, for example where one part brightens as another part fades. This supports Kosch and Scourfield (1992), which found no pattern in the occurrence of various modes. They rather found that one form could experience both pure intensity fluctuations and pure spatial fluctuations within 20 s of observation, and that parts of a form can show different fluctuating modes. It is however not yet clear if the above suggestions can explain such irregular spatiotemporal characteristics.

In summary, the intensity of the individual auroral patches does not fluctuate with the same increases and decreases across the patch. The patches' shapes, however, are remarkably persistent. This means that the fluctuating auroral patches are incoherent on scale sizes smaller than the individual patch, and at the same time coherent on the scale of the overall patch. This supports the suggestions that there are separate processes controlling the scale size of the whole patch and the subscales inside the patch.

### 5.5. Do Patches Always Drift With $\vec{E} \times \vec{B}$ Velocity?

In a nonrotating reference frame (effectively GSE coordinates) the patches drift with 230–287 m/s in a north-eastward direction, which is what typically could be expected for the convection return flow. However, when compared to the SuperDARN convection velocities (corotating) in Figure 11, they are on average only about 50% of the SuperDARN drifts and not in the same direction. Assuming that the SuperDARN drifts are correct, this indicates that the fluctuating patches do not drift with the $\vec{E} \times \vec{B}$ velocity.

#### 5.5.1. Uncertainty in the SuperDARN Drifts

The SuperDARN drifts can, however, be uncertain. The line-of-sight Doppler velocity estimates by SuperDARN are often found to be smaller than the concurrent ion drifts as measured by LEO satellites (Drayton et al., 2005) and velocity measurements by the EISCAT incoherent scatter radar, likely because the high-frequency waves are scattered by small-scale dense structures with refractive indices well below those that are assumed (Gillies et al., 2010). Another issue that can cause uncertainty are the assumptions in deriving the convection solution. The convection solution (so-called fitted velocities) are partially controlled by the radar line-of-sight measurements where these are available, and empirical data. Our patches are located between radar backscatter at ~4 MLT (see supporting information Movie S6 of the SuperDARN polar plots that indicate the location of the measured backscatter). Regardless, the temporal and spatial resolution of SuperDARN (2 min and 50 km) is borderline for the patches (lifetimes of some minutes and scale sizes of some 10 km). For these reasons we cannot eliminate the possibility that the uncertainty of the SuperDARN drifts can explain that our patches do not drift with the $\vec{E} \times \vec{B}$ velocity.

#### 5.5.2. Patches Drift Relative to Each Other

An example of the complex drift patterns of the patches is that they drift relative to each other. For example, Patches 2 and 4 start out separated by about 40 km and about 8 min later they are adjacent and eventually merge into one patch, suggesting a relative drift speed of roughly 80 m/s, which is comparable to the velocities we measured in the ASI frame of reference corotating with Earth, and thus considerable. Our interpretation of this finding is that the electric field pattern is structured and dynamic and thus not necessarily in conflict with $\vec{E} \times \vec{B}$ drift.

#### 5.5.3. Past Findings

The few studies that have focused on finding the drift of fluctuating aurora argues that it moves with the $\vec{E} \times \vec{B}$ velocity, in possible disagreement with our results. Scourfield et al. (1983) found the average drift velocities of fluctuating auroral forms to be in excellent agreement with the radar velocities for a few minutes, after which they started to deviate in direction of up to 55° (magnitudes do not differ by more than 25%) the following 40 min of the event. Nakamura and Oguti (1987) found the drifts of fluctuating auroral patches and arc fragments from time gradients in position-time diagrams (so-called keograms and ewograms) of all-sky TV data. They then displayed the global drift pattern of the two events and found that they resembled the general ionospheric convection pattern measured by radars. Yang et al. (2015) used time-gradients in ewograms to find the eastward velocity component of five fluctuating patches from three events. They found the eastward patch drifts in the range of 156–550 m/s to be slightly larger but in good agreement with the localized eastward convection velocities from SuperDARN. The northward velocity components are not compared. Further, Yang et al. (2017) used the same technique to identify patch east-west velocities from 357 hrs of fluctuating auroral patches events and found velocities ranging from tens of several hundreds m/s in the corotating frame of reference. They argue that the patches are governed by the convection mainly because they mostly move eastward after midnight and westward before midnight. The conclusions of the above mentioned studies are loosely drawn on basis of the east-west velocity component alone and on the general and statistical global velocity patterns.

#### 5.5.4. Implications

If we assume that the SuperDARN drifts are correct, this indicates that the fluctuating patches drift slower than the $\vec{E} \times \vec{B}$ velocity. This implies that the underlying mechanism also must drift slower than the $\vec{E} \times \vec{B}$ velocity. If the underlying mechanism is controlled by a region of cold plasma at the magnetic equator, this could indicate that the cold plasma does not originate from the ionosphere. If there were to be another source, it would point toward the corotating plasmasphere. The plasmasphere is found to have a rather structured boundary, especially during moderate geomagnetic activity (Sandel et al., 2003), and a plasmaspheric plume has been linked to a subauroral proton arc (Spasojević et al., 2004). However, in the dawn sector auroral oval the occurrence probability of plumes are low (Darrouzet et al., 2008), and the observed patch drifts are found to have a small drift relative to the corotating frame of reference (see Figure 11). Therefore there are no strong alternatives to explain that the underlying mechanism would move differently from the $\vec{E} \times \vec{B}$ velocity.

### 5.6. Implied Inconsistency

Above we concluded, on one hand, that the patch shape is maintained to a very high degree and interpreted this as indicative of the mechanism being located at low altitudes and not in the plasma sheet. On the other hand, we found that the patches do not appear to drift with the SuperDARN-determined $\vec{E} \times \vec{B}$ velocity and concluded that this was in conflict with the frozen-in condition, and thus that the mechanism must be located at higher altitudes. These two findings are clearly in conflict. While the determination of patch shape and drift velocities are highly reliable, it is possible that the SuperDARN drift velocities are uncertain. This would actually be in line with the previous published papers discussed in section 5.5.3, which found that the patches likely drift with the $\vec{E} \times \vec{B}$ velocity.

## 6. Summary and Conclusions

We have provided objective and quantitative measures of the extent to which pulsating auroral patches maintain their shape, drift, and fluctuate in a coherent fashion. We use ground-based all-sky imager observations that provide good spatial and temporal resolution (3.31 Hz) of fluctuating patches that allow for a separation of spatial and temporal variations. We traced four individual fluctuating patches using a manual contouring technique and found the characteristics of shape evolution, within patch coherency and drift of the patches in a nonrotating reference frame. The characteristics of four fluctuating auroral patches from a single event do not allow for general conclusions. Our patches are located in dawn sector at ~4 MLT during substorm activity. On the basis of these four patches we conclude the following:

1. For all of the patches their shape can be considered remarkably persistent with 85–100% of the patch being repeated for 4.5–8.5 min.
2. For the three largest patches the temporal correlation coefficient show a negative dependence on distance. A time-delayed response within all of the patches indicate that the so-called streaming spatiotemporal mode can explain part of the variability. Thus, only one of four patches fluctuates in a coherent fashion.
3. The patches appear to drift differently from the SuperDARN-determined $\vec{E} \times \vec{B}$ convection velocity. However, in a nonrotating reference frame the patches drift with 230–287 m/s in a north-eastward direction, which is what typically could be expected for the convection return flow.
4. The patches drift relative to each other. Our interpretation of this finding is that the electric field pattern is structured and dynamic and thus not necessarily in conflict with $\vec{E} \times \vec{B}$ drift.

Our interpretation of the findings is that the mechanism is located at lower altitudes and not in the plasma sheet and that the patches likely drift with the $\vec{E} \times \vec{B}$ velocity.

Our findings and the findings of Humberset et al. (2016) show that the only parameter that appears to be consistent for fluctuating auroral patches, is their shape. The patches do not fluctuate in a coherent fashion, and the energy deposition is highly variable from one fluctuation to the next. The on-time varies wildly and does not show any correlation to the preceding off-time, nor the peak intensity. This supports the suggestion made by Humberset et al. (2016) that pulsating aurora is a misnomer and that the name fluctuating aurora is more appropriate.

**Acknowledgments**
This study was supported by the Research Council of Norway under contract 223252. I. R. M. is supported by a Discovery grant from Canadian NSERC. The authors acknowledge the use of SuperDARN data, a collection of radars funded by national scientific funding agencies of Australia, Canada, China, France, Japan, South Africa, United Kingdom, and the United States of America. A special thanks to Robin Barnes at the Johns Hopkins University Applied Physics Laboratory for providing the SuperDARN data, which are also freely available from vt.superdarn.org. The authors acknowledge the use of SuperMAG indices and all-sky imager data from the Multi-spectral Observatory of Sensitive EMCCDs (MOOSE, moose.space.swri.edu). The SuperMAG indices were obtained freely from supermag.jhuapl.com. We greatly acknowledge James Weygand for the ACE solar wind data. MOOSE all-sky imager data can be obtained from Robert G. Michell and Marilia Samara. The data analyzed in this study are available in the data set Humberset et al. (2018).